\documentclass[12pt,epsf]{article}
\usepackage{graphicx}
\begin{document}


\def\a{\alpha}
\def\b{\beta}
\def\c{\varepsilon}
\def\d{\delta}
\def\e{\epsilon}
\def\f{\phi}
\def\g{\gamma}
\def\h{\theta}
\def\k{\kappa}
\def\l{\lambda}
\def\m{\mu}
\def\n{\nu}
\def\p{\psi}
\def\q{\partial}
\def\r{\rho}
\def\s{\sigma}
\def\t{\tau}
\def\u{\upsilon}
\def\v{\varphi}
\def\w{\omega}
\def\x{\xi}
\def\y{\eta}
\def\z{\zeta}
\def\D{\Delta}
\def\G{\Gamma}
\def\H{\Theta}
\def\L{\Lambda}
\def\F{\Phi}
\def\P{\Psi}
\def\S{\Sigma}

\def\o{\over}
\def\beq{\begin{eqnarray}}
\def\eeq{\end{eqnarray}}
\newcommand{\gsim}{ \mathop{}_{\textstyle \sim}^{\textstyle >} }
\newcommand{\lsim}{ \mathop{}_{\textstyle \sim}^{\textstyle <} }
\newcommand{\vev}[1]{ \left\langle {#1} \right\rangle }
\newcommand{\bra}[1]{ \langle {#1} | }
\newcommand{\ket}[1]{ | {#1} \rangle }
\newcommand{\EV}{ {\rm eV} }
\newcommand{\KEV}{ {\rm keV} }
\newcommand{\MEV}{ {\rm MeV} }
\newcommand{\GEV}{ {\rm GeV} }
\newcommand{\TEV}{ {\rm TeV} }
\def\diag{\mathop{\rm diag}\nolimits}
\def\Spin{\mathop{\rm Spin}}
\def\SO{\mathop{\rm SO}}
\def\O{\mathop{\rm O}}
\def\SU{\mathop{\rm SU}}
\def\U{\mathop{\rm U}}
\def\Sp{\mathop{\rm Sp}}
\def\SL{\mathop{\rm SL}}
\def\tr{\mathop{\rm tr}}

\def\IJMP{Int.~J.~Mod.~Phys. }
\def\MPL{Mod.~Phys.~Lett. }
\def\NP{Nucl.~Phys. }
\def\PL{Phys.~Lett. }
\def\PR{Phys.~Rev. }
\def\PRL{Phys.~Rev.~Lett. }
\def\PTP{Prog.~Theor.~Phys. }
\def\ZP{Z.~Phys. }

\baselineskip 0.7cm

\begin{titlepage}

\begin{flushright}
UT-06-09
\end{flushright}

\vskip 1.35cm
\begin{center}
{\large \bf
The Polonyi Problem and Upper bound on Inflation Scale in Supergravity
}
\vskip 1.2cm
M. Ibe${}^{1}$, Y. Shinbara${}^{1}$ and T.T. Yanagida${}^{1,2}$
\vskip 0.4cm

${}^1${\it Department of Physics, University of Tokyo,\\
     Tokyo 113-0033, Japan}

${}^2${\it Research Center for the Early Universe, University of Tokyo,\\
     Tokyo 113-0033, Japan}

\vskip 1.5cm

\abstract{
We reconsider the Polonyi problem in gravity-mediation models for supersymmetry 
(SUSY) breaking.
It has been argued that there is no problem in the dynamical SUSY breaking scenarios, 
since the Polonyi field acquires a sufficiently large mass of the order of the dynamical 
SUSY-breaking scale $\L_{\rm SUSY}$.
However, we find that a linear term of the Polonyi field in the K\"ahler potential 
brings us back to the Polonyi problem, unless the inflation scale is sufficiently low, 
$H_{\rm inf}\lsim 10^{8}$\,GeV, or the reheating temperature is extremely low, $T_{R}\lsim 100$\,GeV.
Here, this Polonyi problem is more serious than the original one, since
the Polonyi field mainly decays into a pair of gravitinos.
 }
\end{center}
\end{titlepage}

\setcounter{page}{2}

\section{Introduction}

In gravity-mediation models for supersymmetry (SUSY) breaking, gaugino
masses in the SUSY standard model (SSM) are given by a singlet field
$F$-term in a hidden sector. This
singlet field $S$ called as Polonyi field should be an elementary field,
since the gaugino masses are suppressed by higher powers of the Planck 
scale $M_{G} \simeq 2.4\times 10^{18}$ GeV, otherwise. 
This singlet field is completely neutral of any symmetry and the origin of the field
has no enhanced symmetry. 
Thus, the minima  of its potential during inflation and at the true vacuum are
different from each other. The distance between those minima is likely of the order 
of the Planck scale, ${\mit \Delta} S\simeq {\cal O}(M_{G})$. Therefore, 
after the inflation the Polonyi $S$ field starts a coherent oscillation 
around the true minimum when the Hubble parameter becomes of the order of 
the mass of $S$ and its energy density dominates the early universe if
its lifetime is much longer than that of the inflaton. 
As long as 
there is no physical scale besides the Planck scale at high energies, 
the Polonyi field has a mass of the order of the gravitino mass $m_{3/2}={\cal O}(1)$\,TeV
 and it decays to the SSM particles at very late times. 
 The energetic photons and
hadrons produced by the decay deconstruct the light nuclei created by
the big-bang nucleosynthesis (BBN). This is called ``the Polonyi problem''~\cite{Polonyi}.

Possible solutions to the above problem may be found if one introduces a
new high energy scale $M_*$ besides the $M_{G}\, (M_* < M_{G})$. 
There are two possibilities to use this new scale. 
One is to increase interactions between hidden and observed sectors to make 
the decay of $S$ faster,  for instance,  $K= (1/M_*) S^{\dagger}q^{\dagger}q + h.c.$
in a K\"ahler potential. 
Here, $q$ denotes quarks in the SSM.
However, such interactions also increase the gaugino masses and one should 
decrease the gravitino mass by a factor $M_*/M_{G}$ to keep the gaugino masses at the
order of $1$\,TeV. 
Then, the Polonyi mass is reduced also by the same factor, which results in even worse situation. 
The second possibility is to increase the interactions among fields in the hidden sector. 
Namely, one introduces $K= (1/M^2_*)(S^{\dagger}S)^2$ for instance.
Then, the mass of the Polonyi field becomes larger than the gravitino mass $m_{3/2}$ 
and the $S$ can decay before the BBN. 

However, even in the second possibility, there arises another kind of Polonyi problem.
The Polonyi field is much heavier than the gravitino and hence it decays mainly into a pair of gravitinos
which, in turn, results in ``the gravitino overproduction problem''~\cite{kkm}.
Thus, there is still a severe upper bound on the relic abundance of the Polonyi field.
(Notice that, even if one increases the interactions with the SSM particles as above, 
as well as the mass of the Polonyi field, the decay into gravitino is 
still a dominant decay mode of the Polonyi field.)

Despite the above gravitino overproduction problem, the second possibility is very interesting. 
The presence of the new cut-off scale $M_*$ will stop $S$ to run away from the origin
and the distance
between the potential-minimum points of $S$ during the inflation and
$S$ at the true vacuum becomes of the order of $M_*$, that is
$\mit \Delta S\simeq {\cal O}(M_{*})$.
Hence, the coherent oscillation of the Polonyi field $S$ does not necessarily
dominate the universe, if $M_{*}$   is sufficiently smaller than the Planck scale $M_{G}$.
Thus, as shown in the next section, the gravitino overproduction problem from the Polonyi decay can be
evaded for small values of the new scale $M_{*}$ as $M_* \lsim 10^{12-13}$\,GeV.
This result strongly suggests that 
the new scale is nothing but the dynamical scale of the SUSY breaking.

The purpose of  this  letter is to examine if the Polonyi problem
is solved when SUSY is dynamically broken by some strong gauge interactions.
In the above argument, we have assumed $\mit\D S\simeq {\cal O}(M_{*})$.
However, we find that it is not always the case, 
since the potential of the $S$ is
flatter than the mass term $m_{s}^{2} |S|^{2}$ above $S\simeq M_{*}$.
Thus, a careful analysis on the potential of $S$ is required during the
inflation. 
We show as a result that there is a stringent constraint on the Hubble parameter of the inflation or on 
the reheating temperature for a successful solution, that is, $H_{\rm inf} \lsim 10^{8}$\,GeV or
$T_{R} \lsim 10^{2}$\,GeV. 
This concludes that the gravity-mediation models favor relatively low-energy scale inflations
as realized naturally in new inflation models. 
We consider that the present conclusion is quite generic, 
although we derive it in a class of SUSY-breaking models.

\section{Upper bound on the new scale $M_{*}$}

Before going to discuss the dynamical SUSY-breaking models, we show that
there is an upper bound on the new energy scale $M_{*}$.
As we have discussed in Introduction, we assume, for a moment, that 
the presence of the new scale $M_*$ may 
set the distance between minimum points of $S$ during the inflation and $S$ at the 
true vacuum to the order of $M_*$, that is $\mit\Delta S\simeq M_{*}$.

After the inflation, the value of the Polonyi field is fixed at the $\mit\Delta S \simeq M_{*}$ until
the Hubble parameter $H$ falls to the mass of the Polonyi field, $H\simeq m_{s}$, and
then, the Polonyi field begins a coherent oscillation around its true minimum.
Here the mass of the Polonyi field is enhanced by a factor of $1/M_{*}$ compared to the gravitino
mass as discussed in Introduction,%
\footnote{Here and henceforth, we have taken the unit with the
reduced Planck scale, $M_{G}=1$.}
\begin{eqnarray}
m_{s} \simeq \frac{m_{3/2}}{M_{*}}.
\label{eq:massZ}
\end{eqnarray}
The Polonyi field $S$ and the inflaton $\phi$ decay into radiation 
when the Hubble parameter becomes at the decay rate of the Polonyi field, 
\begin{eqnarray}
\Gamma_{s} \simeq \frac{3}{288\pi} \frac{m_{s}^{5}}{m_{3/2}^{2}},
\label{eq:gammaZ}
\end{eqnarray}
and at that of the inflaton,
\begin{eqnarray}
\Gamma_{\phi} = \left(\frac{\pi^{2} g_{*}}{90}\right)^{1/2} T_{R}^{2},
\label{eq:gammaSM}
\end{eqnarray}
respectively.
In Eq.~(\ref{eq:gammaZ}), we present a decay rate of the Polonyi field
into a pair of gravitinos, since the Polonyi field mainly decays into a gravitino
pair.%
\footnote{Recently, it has been extensively discussed that 
the decay into a pair of  gravitinos of the moduli fields~\cite{Moduli} and the inflaton~\cite{Inflaton} 
is much enhanced by even small mixings of those fields with the Polonyi field.
}
In Eq.~(\ref{eq:gammaSM}), we have parametrized the decay rate of the inflaton by using 
a reheating temperature $T_{R}$ after the inflation, and $g_{*}\simeq 200$
denotes  the effective massless degrees of freedom of the SSM.
In the following discussion, we assume that the Polonyi field decays fast enough not to
dominate the energy density of the universe after the inflaton decay.

The ratio of the number density of the gravitino to entropy is given by (after the inflaton decay),
\begin{eqnarray}
Y_{3/2}=\frac{n_{3/2}}{s} \gsim 2\frac{3 T_{R}}{4 m_{\phi}} \frac{n_{s}}{n_{\phi}} { B_{R}}.
\label{eq:yield}
\end{eqnarray}
Here, we have assumed that most of the gravitinos are produced by the Polonyi decay.
$n_{s}$ and $n_{\phi}$ denote the number densities of the Polonyi field and the inflaton
at $H \simeq m_{s}$, $B_{R}={\cal O}(1)$ the branching ratio of the Polonyi decay
into a pair of gravitinos, and $m_{\phi}$ the mass of the inflaton.
The factor $3T_{R}/4 m_{\phi}$ comes from the dilution of the gravitino by the entropy
production of the inflaton decay.
(The equality in Eq.~(\ref{eq:yield})
holds as long as the Polonyi field $S$ decays before its domination of the 
energy density.)
The number density of the Polonyi field and the inflaton at $H\simeq m_{s}$ are given by
(when the Polonyi field starts the oscillation),
\begin{eqnarray}
  n_{s} &\simeq& m_{s} | {\mit\Delta} S|^{2},\\
  n_{\phi} &\simeq& \frac{\rho_{\phi}}{m_{\phi}} \simeq \frac{3 H^{2}}{m_{\phi}}.
\end{eqnarray}
Thus, the yield of  the gravitino can be expressed by,
\begin{eqnarray}
Y_{3/2} \gsim \frac{T_{R}}{2 m_{s}} \left({\mit \Delta} S \right)^{2} B_{R}
\simeq \frac{T_{R}}{2 m_{3/2}} M_{*}^{3} B_{R},
\label{eq:yield2}
\end{eqnarray}
where we have used $\rho_{\phi} = 3 H^{2}\simeq 3 m_{s}^{2}$ and  Eq.~(\ref{eq:massZ}).

To keep the success of the BBN, the gravitino abundance must satisfy~\cite{kkm},
\begin{eqnarray}
Y_{3/2}\, \lsim\, 10^{-(14-16)},
\label{eq:BBN}
\end{eqnarray}
for $m_{3/2} = {\cal O}(1)$\,TeV.
>From Eq.~(\ref{eq:yield2}), we  find that an upper bound on the new scale $M_{*}$, 
\begin{eqnarray}
 M_{*}\lsim 10^{12-13} \,{\rm GeV} \left(\frac{m_{3/2}}{1\,{\rm TeV}}\right)^{1/3} 
 \left( \frac{10^{6}\, {\rm GeV}}{T_{R}}\right)^{1/3}{B_{R}}^{-1/3}.
 \label{eq:Mstar}
\end{eqnarray}
Notably, the above upper bound on $M_{*}$  is close to the scale of  the dynamical SUSY breaking,
\begin{eqnarray}
 \Lambda_{\rm SUSY} \simeq \sqrt{3^{1/2}m_{3/2}}  \simeq  10^{11}\,{\rm GeV}
 \left(\frac{m_{3/2}}{1\,{\rm TeV}}\right)^{1/2}.
 \label{eq:susy}
\end{eqnarray}
Therefore, the above constraint (\ref{eq:Mstar}) strongly suggests that
the new scale $M_{*}$ is nothing but the dynamical scale $\L \simeq 4\pi \L_{\rm SUSY}$
of strong interactions for the SUSY breaking, where
the Polonyi field $S$ obtains its mass 
$m_{s}\sim m_{3/2}/{\L}\simeq \L_{\rm SUSY}$.

\section{Upper bound on the Inflation scale}
As we have seen in the previous section, the solution to the Polonyi problem using 
the new cut-off scale $M_{*}$ suggests the dynamical SUSY breaking by strong interactions.
In this section, we discuss the Polonyi problem to examine if the dynamical SUSY breaking model can indeed solve the problem.
As we have warned in Introduction, we cannot apply the result of the previous section directly,
since the potential of $S$ is very flat above $S\simeq M_{*}$ and our assumption
 ${\mit\Delta} S \simeq M_{*}$ is not guaranteed automatically.
 Thus, we have to arrange the inflaton potential to keep  ${\mit\Delta} S \lsim M_{*} (\simeq \L)$
 during the inflation.
 We will show in this section that there is a stringent upper bound on the inflation scale.

\subsection{The scalar potential of a flat direction}
Before going to discuss the dynamics of the Polonyi field $S$ during the inflation,
we consider the scalar potential of the Polonyi field in  the dynamical
SUSY breaking model~\cite{Izawa:1995jg,IY}.
To see it explicitly, we adopt a dynamical SUSY breaking model based on the 
SUSY $SU(2)$ gauge theory with four fundamental fields $Q_{i}\,(i=1-4)$ and six singlet fields
$S_{ij}=-S_{ji}\,(i,j=1-4)$. 
The tree-level superpotential is given by~\cite{IY},
\begin{eqnarray}
W = \frac{\l_{ij}}{2} S_{ij} Q_{i}Q_{j}.
\label{eq:super1}
\end{eqnarray}
Here, $\l_{ij}$'s denote coupling constants and we have omitted the gauge indices and the 
summations over $i,j$. 
The equations of motion for $S_{ij}$, $\partial W/\partial S_{ij}=0$, set $Q_{i}Q_{j}=0$,
which contradict with the quantum modified constraint Pf$(Q_{i}Q_{j})=\L^{4}$~\cite{Intriligator:1994jr}.
Here, $\L$ denotes the dynamical scale of the gauge interactions.
Hence, the SUSY is broken dynamically.

In this model, there is 
 a  flat direction which is a linear combination $S$ of the 
singlets, $S_{ij}$, which corresponds to the Polonyi field in the previous section.
For $S$ near the origin, $S\ll \L/\l$, the superpotential Eq.~(\ref{eq:super1}) can be effectively
written as,
\begin{eqnarray}
W_{eff}  \simeq \l \left(\frac{\L}{4 \pi} \right)^{2} S,
\label{eq:super2}
\end{eqnarray}
by means of the quantum constraint Pf$(QQ)=\L^{4}$.
Here, we have used a naive dimensional counting~\cite{NDC},
and $\l$ denotes an appropriate linear combination of $\l_{ij}$.
On the contrary, for large values of $S$, 
$S\gg \L/\l$, the $Q_{i}$'s become massive and can be integrated out.
Thus, the theory exhibits a gaugino condensation which produces an effective superpotential,
\begin{eqnarray}
W_{eff}  \simeq \l \left(\frac{\L}{4 \pi} \right)^{2} S.
\label{eq:super3}
\end{eqnarray}
Therefore, the scalar potential for all range of $S$ is given by,
\begin{eqnarray}
V(S) = \left| \l \left(\frac{\L}{4 \pi} \right)^{2}\right|^{2},
\end{eqnarray}
and $S$ is a flat direction.
 
The degeneracy of the above flat direction is lifted by quantum-effects 
in the K\"ahler potential~\cite{Izawa:1995jg}.
For small value of $S$ ($S \ll \L/\l$) the effective K\"ahler potential is expected to take a form,
\begin{eqnarray}
K = |S|^{2}+\frac{\L^{2}}{16\pi^{2}}\left( -\frac{\eta}{4}\left|\frac{\l S}{\L}\right|^{4}
+\cdots
\right),
 \label{eq:effKah}
\end{eqnarray}
where $\eta$ is a real constant which we expect to be of order one, and 
hereafter, $\l$ denotes the coupling constant at  the dynamical scale $\L$.
It leads to a mass term of $S$ as,
\begin{eqnarray}
V_{\rm loop} \simeq \left| \l \left(\frac{\L}{4 \pi} \right)^{2}\right|^{2}\left[1
+ \frac{\eta|\l|^{4}} {(4\pi\L)^{2}} |S|^{2}\right].
\label{eq:massterm}
\end{eqnarray}
We find the mass of the Polonyi field as
\begin{eqnarray}
  m_{s}^{2} \simeq \frac{\eta|\l|^{4}}{(4\pi\L)^{2}}\left| \l \left(\frac{\L}{4 \pi} \right)^{2}\right|^{2}
 \simeq  \sqrt{3} \eta\l
 \left(\frac{\l}{4\pi}\right)^{4}
 m_{3/2}\gg m_{3/2}^{2},
   \label{eq:dynmz}
\end{eqnarray}
where we have used a definition, 
\begin{eqnarray}
m_{3/2} = \frac{1}{\sqrt 3}\left| \l \left(\frac{\L}{4 \pi} \right)^{2}\right|.
\label{eq:gravitino}
\end{eqnarray}

Note that for $\eta >0$, the mass squared of $S$ becomes positive and the vacuum expectation value (VEV) is $\vev{S}=0$.
On the contrary, for $\eta<0$ the mass squared of $S$ is negative and we expect 
$\vev{S}\sim \L/\l$, since the effective potential is lifted in the large $S$ region 
(see Eq.~(\ref{eq:logterm}))~\cite{Arkani-Hamed:1997ut}.
In the following, we only consider the case of $\eta>0$ and $\vev{S}=0$ for simplicity,
since the following discussion will not be changed significantly for $\eta <0$.%
\footnote{For small values of $\l$, $|\l| \lsim {\cal O}(4\pi)$,  the mass squared of the Polonyi field 
is dominated by calculable one-loop corrections and $m_{s}^{2}$ is shown to be 
positive~\cite{Chacko:1998si}.}

For large values of $S$ ($S\gg \L/\l$),
quantum corrections to the scalar potential come from the perturbative wave
function renormalization factor of $S$
and the potential is given by~\cite{Arkani-Hamed:1997ut} ,
\begin{eqnarray}
 V_{\rm loop} 
 \simeq \left| \l \left(\frac{\L}{4 \pi} \right)^{2}\right|^{2} 
 \bigg[ 1 - \int^{\ln \l S}_{\ln \L} \frac{\l(\m)^{2} }{ 4 \pi^{2} } d\ln\m \bigg]^{-1},
 \label{eq:logterm}
\end{eqnarray}
where $\m$ denotes the scale of the renormalization group%
\footnote{Here, we are assuming that $\l_{ij}\sim \l$.}
Thus, the scalar potential in Eq.~(\ref{eq:logterm}) is much flatter than 
Eq.~(\ref{eq:massterm}).
The flatness of the potential for the large field value is a generic feature of 
any effective O'Raifeartaigh models where flat directions are lifted by 
the quantum corrections.

Before closing this subsection, we stress an important feature of the Polonyi field 
in gravity-mediation models.
In gravity-mediation models, the gauginos in the SSM  obtain the SUSY breaking masses
via direct couplings to the Polonyi field,
\begin{eqnarray}
 W = \frac{S}{M_{G}} {\cal W}^{\a}{\cal W}_{\a},
   \label{eq:gaugino}
\end{eqnarray}
 where ${\cal W}^{\a}$'s denote  gauge field strength chiral superfields.
Hence, we expect that the Polonyi field must be neutral under any symmetries.
This means that, in general, we cannot forbid a linear term of the Polonyi field in the K\"ahler 
potential, 
 \begin{eqnarray}
  K  = |S|^{2} -  c^{*} S - c S^{\dagger} + \cdots,
  \label{eq:Kahler}
\end{eqnarray}
where $c $ is a dimensionful parameter and is expected to be of the order of the Planck
scale $M_{G}=1$.
Furthermore, even if we set $c = 0$ at the tree-level, the interaction terms such as 
Eq.~(\ref{eq:gaugino}) generate the linear term of order $c = {\cal O}(N_{g}/16\pi^{2})$ 
at one loop level, where $N_{g}$ is the number of the gauge multiplets circulating
in the loop diagrams.   
Thus,  we naturally expect that the linear term is at least of order $10^{-2}$, i.e. $|c| \gsim 10^{-2}$. 
As we see in the following discussion the linear term in the K\"ahler potential 
has a serious effect on
the dynamics of the Polonyi field during the inflation.

\subsection{Effects of the Hubble parameter during inflation}
Now, let us consider the dynamics of the Polonyi field during the inflation.
First, we assume that the Polonyi field is set to the origin $S = 0$ at the beginning of the inflation.
We do not discuss, here, what physics  provides such a desired situation,
since it is beyond the scope of this letter.
(If the $Q_{i}$'s are in the thermal bath before the inflation, the Polonyi field $S$ acquires
the thermal mass which drives the $S$ to the origin $S = 0$.
This may  be a possible candidate for the physics.)

Once the inflation starts, the effective potential of the Polonyi field and the inflaton $\phi$ are given by,%
\footnote{We can easily extend the following discussion to inflation models
with many fields.}
\begin{eqnarray}
 V &=& e^{K} (K_{\bar{I}J}^{-1}(D_{I}W)^{\dagger} (D_{J} W) -3 |W|^{2} ),\,\, (I = \phi, S),
 \label{eq:potential}
\end{eqnarray}
where we have assumed the K\"ahler potential as,
\begin{eqnarray}
  K = |\phi|^{2}  +  |S|^{2} - c^{*} S  - c S^{\dagger},
\end{eqnarray}
for simplicity.
$D_{I}W$ and $K_{\bar{I}J}$ are defined by
\begin{eqnarray}
D_{I} W &=& \frac{\partial W}{\partial X_{I}} + \frac{\partial K}{\partial X_{I}} W,\,\, (X_{I} = \phi, S),\\
K_{\bar{I}J} &=& \frac{\partial^{2} K}{\partial X_{I}^{\dagger} X_{J}},\,\, (X_{I} = \phi, S).
\end{eqnarray}
In addition, we also assume  that the hidden sector and the inflaton sector is separated in
the superpotential $W$ as,
\begin{eqnarray}
W = W(S)+W(\phi).
\end{eqnarray}
By using  $V_{\rm loop}$ a potential of $S$ given in Eqs.~(\ref{eq:massterm}) and (\ref{eq:logterm}),
$V_{\rm inf}$ a potential of the inflaton which is nearly constant during the inflation, and $K(S)$ 
a K\"ahler potential for the Polonyi field $S$, the above scalar potential can be rewritten as,
\begin{eqnarray}
V  \simeq& e^{K(S)} (V_{\rm inf} + V_{\rm loop}(S)).
\end{eqnarray}
During the inflation, the potential of the Polonyi field 
is changed from the one at the true vacuum due to the first term, $e^{K(S)}V_{\rm inf}$.
Here, we have neglected $K(\phi)$, since it is irrelevant to our discussion. 

As we have mentioned at the end of the previous subsection, we cannot forbid the linear term in the 
K\"ahler potential.
During the inflation, such a linear term leads to a slope of the Polonyi potential around the true vacuum
$S=0$,
\begin{eqnarray}
 V =  e^{K(S)}V_{\rm inf} \simeq -(c^{*}S + c S^{*}) V_{\rm inf}.
 \label{eq:linear}
\end{eqnarray}
On the other hand, the Polonyi potential in Eq.~(\ref{eq:logterm}) is nearly 
flat for $S\gsim \L/\l$ with a height of
\begin{eqnarray}
V(S) = V(S\gg \L) = \xi \left| \l \left(\frac{\L}{4 \pi} \right)^{2}\right|^{2}
= 3 \xi m_{3/2}^{2},
\label{eq:flat}
\end{eqnarray}
where, $\xi$ is a numerical constant at most of order one.
Thus, if the linear term in Eq.~(\ref{eq:linear}) is large, the potential minimum is shifted from
the origin $S=0$.

To investigate the behavior of $S$ during the inflation more closely, we approximate the above
 Polonyi potential Eq.~(\ref{eq:massterm}) and (\ref{eq:flat}) 
by,
\begin{eqnarray}
V_{\rm loop}(S) =
\left\{
\begin{array}{lr}
\displaystyle{m_{s}^{2} |S|^{2}},  & (|S| \leq |S_{*}|),   \\
3 \xi m_{3/2}^{2}, &  (|S|> |S_{*}|).  
\end{array}
\right.
\label{eq:app}
\end{eqnarray}
Here, $m_{s}$ is the Polonyi mass given in Eq.~(\ref{eq:dynmz}) and 
$S_{*}$ is a field value where higher order terms of the effective K\"ahler potential in 
Eq.~(\ref{eq:effKah}) become important.
In the following analysis, we take
\begin{eqnarray}
S_{*} =  \eta'\frac{\L}{\l},
\label{eq:upper}
\end{eqnarray}
with $\eta'={\cal O}(1)$.

\begin{figure}[t!]
\begin{center}
\begin{minipage}{0.325\hsize}
\includegraphics[width = 1 \linewidth]{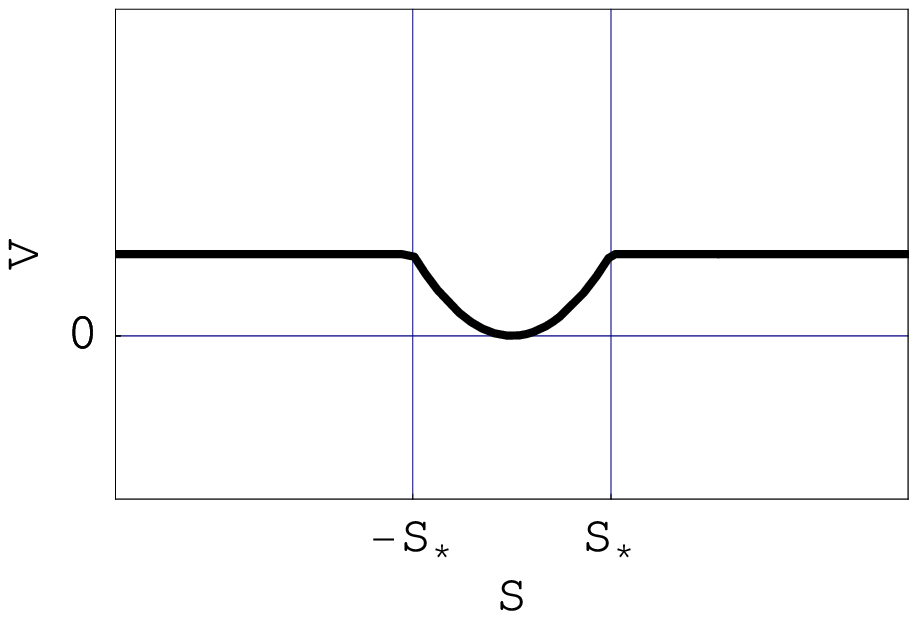}
\end{minipage}
\begin{minipage}{0.325\hsize}
\includegraphics[width = 1 \linewidth]{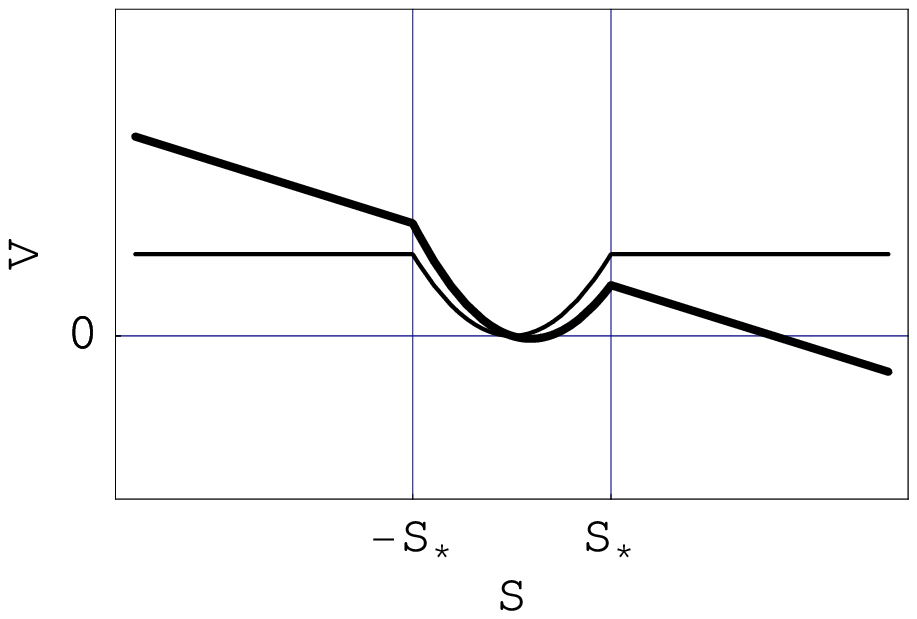}
\end{minipage}
\begin{minipage}{0.325\hsize}
\includegraphics[width = 1 \linewidth]{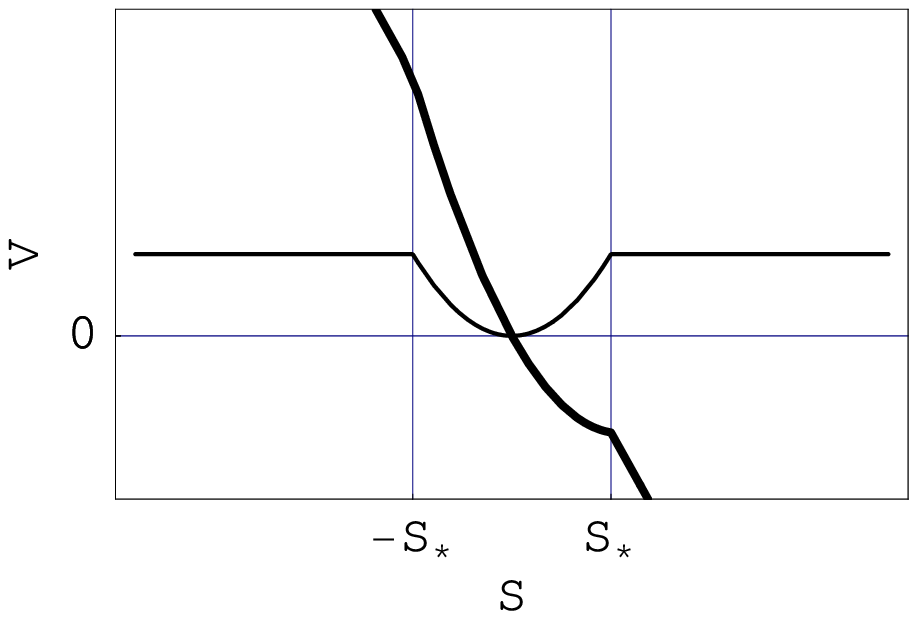}
\end{minipage}
\caption{ Schematic plots of the Polonyi potential during the inflation.
Here, we assume that $\l = 4\pi$, $\xi = \eta' =1$, $c\simeq M_{G}$ and $m_{3/2}=1$\,TeV.
The potentials correspond to $H_{\rm inf} = 0$,  $5\times 10^{6}\,{\rm GeV}$, 
$10^{8}\,{\rm GeV}$ from left to right, respectively.}
\label{fig:potential}
\end{center}
\end{figure}

Under the above approximation, we find that the true minimum ($S= 0$) is substantially shifted
unless a condition,
\begin{eqnarray}
(c^{\dagger} S_{*}+c S_{*}^{\dagger}) V_{\rm inf}\, \lsim 
3 \xi m_{3/2}^{2},
\label{eq:broken}
\end{eqnarray}
is satisfied.
By using Eqs.~(\ref{eq:gravitino}) and (\ref{eq:upper}),
this condition can be expressed as an upper bound on
 the Hubble parameter $H_{\rm inf}$ during the inflation,
\begin{eqnarray}
 H_{\rm inf}\lsim 10^{8}\, {\rm GeV }\times \left(\frac{m_{3/2}}{\rm TeV}\right)^{3/4} 
 \left(\frac{10^{16}\,{\rm GeV} }{|c|}\right)^{1/2}
  \left(\frac{\l}{4\pi}\right)^{3/4}
 (\xi\eta'^{-1})^{1/2},
 \label{eq:Hubble}
\end{eqnarray}
where we also used the relation $3H_{\rm inf}^{2} = V_{\rm inf}$.%
\footnote{We assume  inflation models with $F$-term potentials in this letter.
However, if we consider $D$-term inflation models, it depends on details of 
the models if we may evade the constraint Eq.~(\ref{eq:Hubble}).}
Figure.~\ref{fig:potential} is a schematic figure of the Polonyi potential around $S= 0$
for various values of $H_{\rm inf}$.
The figure shows that the minimum at $S= 0$ is substantially shifted from the origin when 
the above condition is violated.

Notice that we have derived an upper bound on the Hubble parameter Eq.~(\ref{eq:Hubble})
based on a specific model of the dynamical SUSY-breaking.
However, by using a naive dimensional counting~\cite{NDC}, 
we can approximate Polonyi potentials by Eq.~(\ref{eq:app}) with $m_{s}\lsim \L$
and $S_{*}\gsim \L/4\pi$ for any effective O'Raifeartaigh models where flat directions
are lifted by the quantum corrections.
Thus, we consider that the above upper bound on the Hubble parameter
is a quite generic result for any effective O'Raifeartaigh models.

Once the minimum at $S=0$ disappears during the inflation, $S$ rolls down to $c$ immediately.
In such a case, we suffer from the recurrence of the Polonyi problem, 
since $S$ is fixed at $S \simeq c$ until the Hubble parameter becomes very small.
As we see in the next section, the spilled Polonyi field causes a gravitino
overproduction problem for $|c|\gsim 10^{-2}$, 
if $T_{R}\gsim 1-100$\,GeV (see Eq.~(\ref{eq:TRbound})).
Thus, for $|c|\gsim 10^{-2}$, the above upper bound on the Hubble parameter, 
$H_{\rm inf}\lsim 10^{8}$\,GeV, 
is a necessary condition for the inflation in the gravity-mediation models,
unless the reheating temperature is extremely low, $T_{R}\lsim100$\,GeV.%
\footnote{For the leptogenesis~\cite{lepto} to work, we need 
a reheating temperature $T_{R}$ higher than the critical temperature for the 
electroweak phase transition~\cite{lepto,Ibe:2005jf}.
Besides, it seems rather difficult for the inflation to achieve the low reheating temperature such as $T_{R}\ll 10^{6}$\,GeV.}

This result shows that the SUSY chaotic inflation~\cite{Chaotic}  (where typical Hubble parameter is 
$H_{\rm inf} \simeq 10^{14}$\,GeV)
and the SUSY topological inflation~\cite{Topological}  ($H_{\rm inf}\simeq 10^{11-14}$\,GeV) are  disfavored.
The SUSY hybrid inflations are also disfavored since the typical Hubble parameters
are $H_{\rm inf} \simeq 10^{13-15}$\, GeV~\cite{Hybrid}.

On the contrary, among inflation models constructed in SUGRA, 
a new inflation model in~\cite{New},
is one of the most attractive candidates.
The model has a flat inflaton potential,
\begin{eqnarray}
 V(\varphi) \simeq v^4 - \frac{k}{2}v^4 \varphi^2
 -\frac{g}{2^{\frac{n}{2}-1}}v^2\varphi^n
 +\frac{g^2}{2^n}\varphi^{2n}, \, (n\geq 3),
\label{eq:newinf}
\end{eqnarray}
where $v$ is the energy scale of the inflation, $g$ the coupling constant in the 
superpotential, and $k$ is the quartic coupling constant in the K\"ahler potential.
>From the COBE normalization, the inflation scales are determined for $k\lsim 10^{-2}$~\cite{IS},
\begin{eqnarray}
H_{\rm inf} &=& \mbox{undetermined by the COBE normalization}, (n=3),\\
H_{\rm inf} &\simeq& 10^{5.4}\,{\rm GeV} \times \frac{1}{g},\, (n=4),\\
H_{\rm inf} &\simeq& 10^{8.6}\,{\rm GeV} \times \frac{1}{g^{1/2}},\, (n=5),\\
H_{\rm inf} &\simeq& 10^{9.9}\,{\rm GeV} \times \frac{1}{g^{1/3}},\, (n=6),
\end{eqnarray}
and $H_{\rm inf}$ increases for larger $n$.
Thus, we find that the new inflation model with $n\gsim 5$ is disfavored.
Interestingly, the favored new inflation model of $n=4$ predicts the spectral index 
$n_{s} \simeq 0.94-0.95$~\cite{Ibe:2006fs}
which is well consistent with the recent WMAP result,
 $n_{s}= 0.951^{+0.015}_{-0.019}$ (68\%C.L.)~\cite{WMAP}.%
\footnote{The detailed analysis on the new inflation model of $n=3$ will be
discussed elsewhere~\cite{IS}.
}

\section{Fate of spilled Polonyi field}
As we have seen in the previous section,
if the Hubble parameter is too large and it does not satisfy the condition Eq.~(\ref{eq:Hubble}),
the minimum of the Polonyi potential is shifted away from the origin. 
Then, the Polonyi field falls to the minimum of the potential, $S \simeq c $, during the inflation. 
In this section, we consider the Polonyi problem for such a case.

In general, there are many local minimal points of the Polonyi potential
$V(S)$ around $S={\cal O}(M_{G})$,
and $S$ has a mass of the order of the gravitino mass at each minimal points. 
On the contrary, the curvature of the Polonyi potential Eq.~(\ref{eq:logterm}) 
at $S \simeq c $ is given by,
\begin{eqnarray}
 V''(S) \simeq-\frac{\l(\m_{c})^{2}}{4 \pi^{2} |c|^{2}} \left| \l\left(\frac{\L}{4 \pi} \right)^{2}\right|^{2}
 \simeq -\frac{3\l(\m_{c})^{2} m_{3/2}^{2}}{4 \pi^{2} |c|^{2}},
 \label{eq:curvature}
\end{eqnarray}
with $\m_{c} \simeq \l c$, which is at most the gravitino mass when $c$ is close to $M_{G}$.
Hence, for $c\simeq M_{G}$, we do not expect that the Polonyi field returns to $S= 0$
after the inflation, since the  attractive force towards the origin is weaker than those towards 
the minimal points at $S= {\cal O}(M_{G})$.
In this case, the Polonyi field is attracted to one of the local minimal points at $S={\cal O}(M_{G})$
and starts to oscillate around the local minimal point when the Hubble parameter 
becomes at $H\simeq m_{3/2}$ after the inflation.
Since the typical distance between the point $S\simeq c$ and the local minimal points 
at $S={\cal O}(M_{G})$ is of the order of the Planck scale, such a late time coherent oscillation
causes nothing but the original Polonyi problem~\cite{Polonyi}.
Therefore, once the Polonyi field is spilled out of the $S=0$ minimum, we again suffer 
from the original Polonyi problem for $c\simeq M_{G}$.

On the other hand, if the linear term is somewhat smaller than the Planck scale, 
the Polonyi field is attracted to $S= 0$ after the inflation, since the curvature of 
Eq.~(\ref{eq:curvature}) exceeds the gravitino mass.
Then, the Polonyi field starts to oscillate around its true minimum $S =0$ from $S \simeq c$
when the Hubble parameter $H$ falls to $H_{\rm osc} \simeq \l(\m_{c}) m_{3/2}/c$,%
\footnote{We confine ourselves to the inflation model with the reheating 
temperature $T_{R} \lsim 10^{6-8}$\,GeV,  since otherwise we have the gravitino
 overproduction problem from the scattering process of the thermal background after the
 inflation~\cite{kkm}.
For such reheating temperature, 
we can safely neglect thermal effects to the Polonyi potential from $Q_{i}$ plasma.}
and it eventually decays dominantly into a pair of gravitinos
(see section 2.).

After the whole reheating process, the yield of the gravitino is given by,
\begin{eqnarray}
Y_{3/2} = \frac{n_{3/2}}{s} \gsim 2\frac{3 T_{R}}{4 m_{\phi}} \frac{n_{s}}{n_{\phi}} B_{R},
\label{eq:gravA1}
\end{eqnarray}
where $n_{s}$  and $n_{\phi}$ denote the number densities of the Polonyi field and the inflaton
at $H_{\rm osc}$.
To estimate the yield of the gravitino
 conservatively, we assume that the Polonyi field decays into a pair of
gravitinos with its mass in Eq.~(\ref{eq:dynmz}) immediately after it starts to oscillate around $S=0$.
Then, the number density of the Polonyi field and the inflaton at $H_{\rm osc}$ are given by,
\begin{eqnarray}
 n_{s} &\simeq& \frac{\xi}{m_{s}} \left| \l\left(\frac{\L}{4\pi}\right)^{2}\right|^{2}\simeq \frac{3 \xi m_{3/2}^{2}}{m_{s}},\label{eq:ns}\\
 n_{\phi}&\simeq& \frac{\r_{\phi}}{m_{\phi}}\simeq\frac{3 H_{\rm osc}^{2}}{m_{\phi}}.
\end{eqnarray}
By using these number densities, we obtain the yield,
\begin{eqnarray}
 Y_{3/2} \gsim 2 \frac{3 T_{R}}{4 m_{s}} \xi \l(\m_{c})^{-2} |c|^{2} B_{R}.
\end{eqnarray}
Hence, the BBN constraint~\cite{kkm} in Eq.~(\ref{eq:BBN})  requires,
\begin{eqnarray}
 T_{R} \lsim 1 - 100\,{\rm GeV} 
 \left(\frac{\l}{4\pi}\right)^{5/2}
 \left(\frac{ m_{3/2}}{1\,\rm TeV}\right)^{1/2} \left(\frac{10^{16}\,{\rm GeV}}{|c|}\right)^{2}
 (\xi B_{R})^{-1} \eta^{1/2} \l(\m_{c})^{2}. 
 \label{eq:TRbound}
\end{eqnarray}
Thus, for $|c|\gsim 10^{-2}$, 
we cannot avoid the gravitino overproduction problem even if the spilled Polonyi field 
returns to $S = 0$ after the inflation, unless the reheating temperature is extremely low.%
\footnote{Again, we consider that the above upper bound on $T_{R}$ 
is a quite generic result for effective O'Raifeartaigh models.}

Notice that the above upper bound on $T_{R}$ is conservative since we have used $m_{s}$
in Eq.~(\ref{eq:dynmz}) to estimate the decay rate of the Polonyi field (Eq.~(\ref{eq:gammaZ}))
and the number density of the Polonyi field (Eq.~(\ref{eq:ns})).
Since the amplitude of the Polonyi field is much larger than $\L/\l$ at the beginning 
of the oscillation,  the effective mass of the Polonyi field is smaller than $m_{s}$.
Thus, one may consider that the decay rate of the Polonyi field is smaller 
and $n_{s}$ is larger, which leads to a larger number density of the gravitinos.
However, even if the effective mass of the Polonyi field is much smaller than $m_{s}$,
it behaves as a particle with mass $m_{s}$ during a time period of $1/m_{s}$ in each oscillation.
Thus, the effective decay rate of the Polonyi field (with mass $m_{s}$)
at the beginning of the coherent oscillation is given by,
\begin{eqnarray}
 \G_{s}^{\rm eff} \simeq \frac{H_{\rm ocs}}{m_{s}} \G_{s}.
\end{eqnarray}
As we see from Eqs.~(\ref{eq:gammaZ}) and (\ref{eq:dynmz}), $\G_{s}$ is close to $m_{s}\simeq \L$
for $\l \simeq 4\pi$, and hence, the effective decay rate is comparable to the Hubble 
parameter at the beginning of oscillation, $H_{\rm osc}$.
Thus, the Polonyi field effectively decays with a mass $m_{s}$, immediately after it
starts to oscillate, and the number density of the Polonyi field which
results in the gravitino number density is roughly given by
dividing the energy density in Eq.~(\ref{eq:app})  by $m_{s}$ (see Eq.~(\ref{eq:ns})).
Therefore, we may use safely the above conservative analysis.%
\footnote{
In the model we have considered in section 3, there is a non-anomalous approximate 
$R$ symmetry.
Then, the $Q$-ball and anti-$Q$-ball~\cite{Qball} 
can be formed after $S$ starts to oscillate around $S=0$.
In this case, as long as the annihilation of $Q$-balls can be neglected,
the $Q$-ball has a long lifetime, which increases the resultant gravitino
abundance.
Thus, the upper bound on $T_{R}$ can be much severer than Eq.~(\ref{eq:TRbound})
for the dynamical SUSY breaking model in section 3.
We thank F.~Takahashi for pointing out this.
}

Finally, we comment on possible effects of direct couplings between the 
hidden sector and the inflaton sector in the superpotential.
Although such interactions are highly model dependent (charges of the fields, etc.), 
we at least expect the terms,
\begin{eqnarray}
W =  W(\phi)(1 + c_{1} S + c_{2} S^{2}+\cdots),
\end{eqnarray}
since the Polonyi field is completely neutral under any symmetries.
Such terms, in general, increase the number of the minimal points  around $S ={\cal O}(M_{G})$,
and hence, the above problems are not improved.
\section{Conclusions}
In this letter, we have considered a solution to the Polonyi problem by assuming dynamical
SUSY breaking.
We have found that even for dynamical SUSY breaking models, 
the linear term $c$ of the Polonyi field in the K\"ahler potential may bring us back to the 
Polonyi problem or the gravitino overproduction problem.
To avoid the problems for the most natural case, 
 $|c|\gsim 10^{-2}$, 
 in the gravity-mediation models  the inflation in  the early universe should have
a very small Hubble parameter, $H_{\rm inf}\lsim 10^{8}$\,GeV
(Eq.~(\ref{eq:Hubble})) such as new inflation models, or 
a very low reheating temperature, $T_{R}\lsim 10^{2}$\,GeV 
(Eq.~(\ref{eq:TRbound})).%
\footnote{If there is a late-time entropy production, the constraint on the reheating
temperature can be weakened.}
This result is very interesting since the favored new inflation model in SUGRA naturally
predicts the spectral index as $n_{s}\simeq 0.94-0.95$, which is very consistent with 
the recent WMAP observation.

We comment that there is also no theoretical reason to suppress the K\"ahler interactions such as
$K=\k |\phi|^{2}S^{2}$.
If it exists, the inflaton decay into a pair of gravitinos is enhanced  and 
a stringent constraint on the inflation model is obtained.
The recent analysis has shown that the hybrid inflation model is very disfavored~\cite{Inflaton}.
Furthermore, the interactions between the hidden sector and the inflaton sector in the superpotential
can enhance the inflaton decay rate into a pair of gravitinos, which may give more stringent
constraints.
The detailed analysis in the new inflation 
model~\cite{New} including such superpotential interactions
will be discussed elsewhere~\cite{IS}.

We should note finally that the Polonyi problem discussed in this letter may not exist in 
gauge- or anomaly-mediation models for SUSY breaking.
This is because the Polonyi field may have charges of some symmetries suppressing the 
linear term in the K\"ahler potential or because there is not necessarily  present the elementary
Polonyi field in those models.

\section*{Acknowledgments}
M.~I. thanks the Japan Society for the Promotion of Science for
financial support.  
This work is partially supported by Grand-in-Aid
Scientific Research (s) 14102004.
The work of T.T.Y. has been supported in part by a Humboldt Research Award.

\appendix

\end{document}